\documentclass[showpacs,amsmath,amssymb,twocolumn,floatfix,prl]{revtex4}
\usepackage[dvips]{graphicx}
\input{epsf}

\begin{document}

\title{Treatment of sound on quantum computers}


\author{Jae Weon Lee$^{(a)}$,  Alexei D. Chepelianskii$^{(a,b)}$ and 
Dima L. Shepelyansky$^{(a)}$}
\homepage[]{http://www.quantware.ups-tlse.fr}
\affiliation{$^{(a)}$Laboratoire de Physique Th\'eorique, 
UMR 5152 du CNRS, Univ. P. Sabatier, 31062 Toulouse Cedex 4, France\\
$^{(b)}$Ecole Normale Sup\'erieure, 45, rue d'Ulm, 
75231 Paris Cedex 05, France
}

\date{July 31, 2003}

\begin{abstract}
We study numerically how a sound signal stored in a
quantum computer can be recognized and restored 
with a minimal number of measurements in 
presence of random quantum gate errors.
A method developed uses elements of MP3
sound compression and allows to recover human speech
and sound of complex quantum wavefunctions.
\end{abstract}

\maketitle
In the last decade the rapid technological progress
made possible the treatment of large amounts of information 
and their transmission over large distances. 
In spite of this  the transmission of digital 
audio signals required the development of specific 
compression methods in order to achieve real time
audio communication. 
A well known example of audio compression is the 
Mpeg Audio Layer 3  (MP3) which allows to 
reduce the signal size by an order of magnitude 
without noticeable distortion \cite{mp3}. 
It essentially uses the Fast Fourier Transform (FFT) in 
order to have rapid access to the signal spectrum,
whose analysis allows to reach significant compression rates.
Such methods find every day applications in the
Internet telephone communication and teleconferences. 

Recent developments in quantum information 
make possible a new type of computation 
and communication using the quantum nature of the signal
(see {\it e.g.} \cite{chuang,qcom}).
In quantum computation theory it was shown that certain 
quantum algorithms can be exponentially more efficient than 
any known classical counterpart. For instance 
the Shor algorithm enables to factorize large integers 
in a time polynomial in the number of bits whereas 
all known classical algorithms are exponential \cite{shor}.
This algorithm relies on the Quantum Fourier Transform (QFT) which 
is exponentially faster than FFT \cite{chuang,shor}. 
Simple quantum algorithms 
have been realized experimentally with few qubit quantum 
computers based on Nuclear Magnetic Resonance (NMR) and
ion traps \cite{chuang1,cory,blatt}. Quantum communications
also attracted a great deal of attention since they
allow to realize secret data transmission. Currently, such a
transmission has been achieved over distances of up to a 
few tens kilometers \cite{qcom}.

These prospects rise timely the question 
of treatment of audio signals on quantum computers. 
Classical audio analysis methods cannot be directly applied to 
quantum signals and it is important to adapt them to 
the new environment of quantum computation. 
Furthermore while digital signal treatment is faultless,  
quantum computation contains phase and amplitude errors
which can affect the quality of sounds encoded on a 
quantum computer. In addition the extraction of quantum 
information relies on quantum measurements which bring 
new elements that must be taken into account in the treatment 
of quantum audio signals. Different type of sound signals 
are possible like human speech, music or pure quantum 
objects like the Wigner function \cite{wigner} 
which can be efficiently
prepared on quantum computers \cite{laflamme,levi}.

For the standard audio sampling rate of $44 kHz$ a 
quantum computer with $20$ qubits (two-level quantum 
systems, see \cite{chuang}) can store a mono audio 
signal of $23$ seconds. A quantum computer with
$50$ qubits may store an amount of information 
exceeding all modern supercomputer capacities (1000 years of sound).
Thus the development of readout methods, which 
in presence of imperfections can 
recognize and restore the sound signal via a minimal 
number of quantum measurements, becomes of primary importance.

To study this problem we choose the following
 soundtrack pronounced by HAL 
in the Kubrick movie ``2001: a space odyssey'': 
{\it ``Good afternoon, gentlemen. I am a HAL 9000 computer. 
I became operational at the H.A.L. lab in Urbana, Illinois 
on the 12th of January''} \cite{hal}. The duration of 
this recording is 26 seconds and at a sampling rate
$f = 8 kHz$ it can be encoded 
in the wavefunction of a quantum computer with $n_q = 18$ qubits
(the HAL speech is thus zero padded to last 32 seconds).
This rate gives good sound quality and is more 
appropriate for our numerical studies. 
Digital audio signals can be represented by a sequence of 
samples with values $s_n$ in the interval $(-1, 1)$ so that
the $n$-th sample gives the sound at time 
$t = n / f$. This signal can be encoded on a quantum 
computer by the following wavefunction 
$\psi = A \sum_n s_n | n >$, where $A$ is normalization constant.
The state $|n>$ represents the multiqubit eigenstate 
$|a_1 ... a_i ... a_{n_q}>$ where $a_i$ is 0 or 1 corresponding
to the lower or upper qubit state, the sequence of $a_i$ gives
the binary representation of $n$. 

Using numerical simulations we test various approaches
to the readout problem of the above signal 
encoded in the wavefunction of a quantum computer. 
Direct measurements of the wavefunction do not allow 
to keep track of the sign of $s_n$ and many measurements
are required to determine the amplitude $|s_n|$ with good 
accuracy. Another strategy is to use the analogy 
with the MP3 coding. With this aim we divide the sound 
into consecutive frames of fixed size $\Delta n = 2^{n_f}$
where $n_f$ can be viewed as the number of qubits 
required to store one frame. We choose these qubits 
to be the $n_f$ least significant qubits in the binary 
representation of $n = (a_1 ... a_{n_q-n_f+1} ... a_{n_q})$.
Then we perform QFT on these $n_f$ qubits that corresponds 
to applying FFT to all the  $2^{n_q - n_f}$ frames of the signal 
in parallel. This requires $n_f (n_f + 1) / 2$ quantum gates
contrary to $O(n_f 2^{n_f})$ classical operations for FFT.
After this transformation the wavefunction represents 
the instantaneous spectrum of the sound signal evolving 
in time from one frame to another. 
This way the most significant  $n_q - n_f$ qubits store 
the frame number $k$ while the least significant $n_f$ qubits 
give the frequency harmonic number $j$.
Hence, after QFT the wave function has the form 
$\psi = \sum_{k,j} S_{k,j} |k,j>$ where $S_{k,j}$ is 
the complex amplitude of the $j$-th harmonic in the 
$k$-th frame. The measurements in this representation 
gives the amplitudes $|S_{k,j}|$ while phase information 
is lost. However for sound the main information 
is stored in the spectrum amplitudes and the ear 
can recover the original speech even if the phases 
are all set to zero. Thus the recovered signal is obtained 
by the inverse classical FFT and is given by 
$s'_n = \sum_j |S_{k,j}| e^{ 2 \pi i j m / \Delta n } $
with $n = k \; \Delta n  + m $ (to listen sound we use $Re (s'_n)$).
For time domain measurements of $s_n$ the recovered signal
is $\tilde{s}_n = |s_n|$.
These expressions for $s'_n$ and $\tilde{s}_n$ hold
for an infinite number of measurements.
In reality it is important to approximate them accurately 
with a minimal number of measurements.

\begin{figure}[t!]  
\centerline{\epsfxsize=8.5cm\epsffile{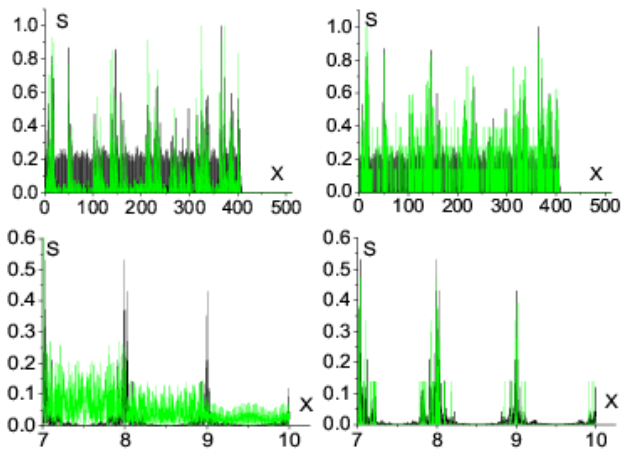}}
\vglue -0.5cm 
\caption{(color online) Sound signal spectrum as a function of frame number 
$x = n / \Delta n $. On all panels the black curves show the spectrum of the 
original signal $s_n$. On the left top panel
the green/gray curve represents the spectrum of $\tilde{s}_n$ 
obtained with $M = 5$ measurements per frame in time domain.
On the right top panel it shows the spectrum of $s'_n$ 
for the same number of measurements $M$ performed after QFT.
Bottom panels show these spectra on a smaller scale.
}
\label{fig1}       
\end{figure}

For our soundtrack we found that the optimal frame size is 
$\Delta n = 2^9 (n_f=9)$
and we perform $M$ measurements per frame.
In Fig. 1 we compare the spectrum of $s'_n$ and $\tilde{s}_n$
with the original signal spectrum. Here only $M = 5$ measurements
per frame are performed and the results clearly show that 
the quality of the restored sound is significantly higher 
for the spectrum domain measurements. 
Examples of restored and original sounds are available at \cite{qaudiosite}.
The HAL speech is recognizable from $M = 5$ for spectrum domain 
measurements while it is distorted beyond recognition 
for direct time domain measurements even for $M = 100$.

\begin{figure}[t!]  
\centerline{\epsfxsize=9.0cm\epsffile{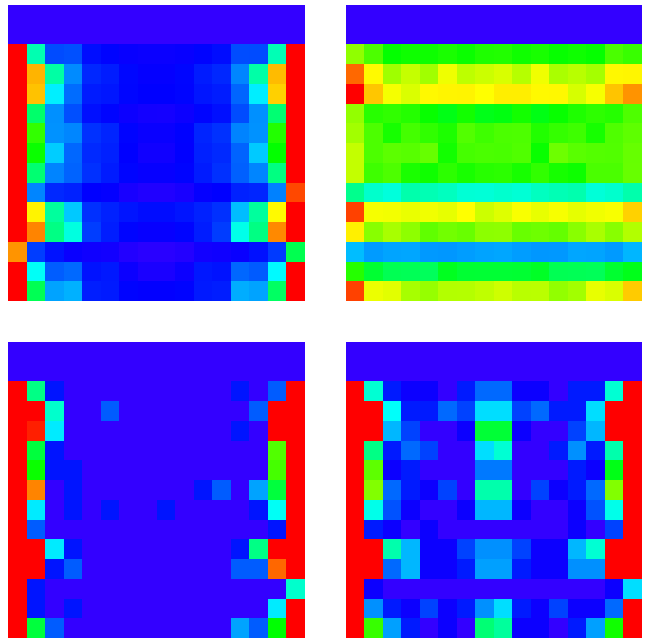}}
\vglue -0.6cm 
\caption{(color) 
Coarse grained diagram for the signal of Fig. 1 obtained 
by measurements of qubits 1 to 4 and 10 to 13 (see text).
Top left: original signal; bottom left: distribution 
obtained after QFT with the same number of measurements as in
Fig. 1; bottom right: same as bottom left with the noise in 
the quantum gates ($\epsilon  = 0.05$); top right: diagram obtained 
from measurements in time domain (with the same number of
measurements as in bottom right).
The color represents amplitude of the spectrum: blue for zero and
red for maximal values.
The horizontal/vertical axis corresponds to coarse grained
frequency/time. 
}
\label{fig2}       
\end{figure}

Fig. 1 shows the global structure of the signal spectrum.
To make comparison more quantitative and visual we show 
coarse grained color diagrams of the spectrum.
The coarse graining is obtained by measuring 
only certain qubits corresponding for example to 
$a_1 a_2 a_3 a_4$ and $a_{10} a_{11} a_{12} a_{13}$. 
For $s'_n$ this gives a coarse grained spectrum $|S_{k,j}|$
in $2^4 \times 2^4$ cells
shown in Fig. 2. The same total number of measurements 
as in Fig. 1 allows to reproduce the original 
coarse grained diagram with good accuracy. 
Even if QFT is performed with noisy gates 
(the angle in the unitary rotations fluctuates with
an amplitude $\epsilon \pi = 0.05 \pi$)
the spectrum diagram remains stable and is 
reproduced with good accuracy (see Fig. 2).
At the same time the coarse graining in the time domain 
for the signal $\tilde{s}_n$ 
shown in Fig. 1 gives the diagram which is very different from the 
original.

\begin{figure}[t!]
\epsfxsize=3.2in
\epsfysize=2.6in
\epsffile{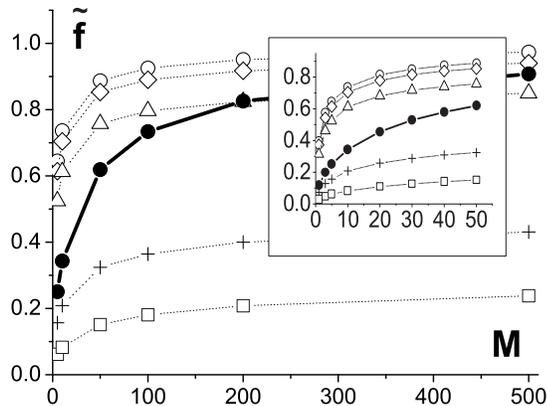}
\vglue -0.7cm
\caption{Fidelity $\tilde{f}$
for sound signals $s'_n$ (open circles)
and $\tilde{s}_n$ (full circles) 
as a function of number of measurements
$M$. For $s'_n$ fidelity is shown for
various amplitudes of noise in the QFT gates with $\epsilon = 0$ (o),
0.05 (diamonds), 0.1 (triangles), 0.3 (+), 1 (squares).
Inset shows data at small $M$ scale.
}
\label{fig3}
\end{figure}

The global quality of the recovered signal
$\tilde{s}_n$ (or $s'_n$) obtained via a finite number of measurements
is convenient to characterize by the fidelity
defined as $\tilde{f} = 
|\sum_n{ \tilde{s}^{(')*}_n \tilde{s}^{(')}_n(M,\epsilon)}|/R$ with 
$R= 
(\sum_n {|\tilde{s}^{(')}_n|^2} 
\sum_n {|\tilde{s}^{(')}_n(M,\epsilon)|^2})^{1/2}$. 
Here, $\tilde{s}^{(')}_n(M,\epsilon)$
is the signal obtained in a way described above with $M$
measurements per frame in time domain $(\tilde{s}_n(M))$
or in frequency domain after QFT with noisy gates $(s'_n(M,\epsilon))$.
The dependence of $\tilde{f}$ on $M$ is shown in Fig. 3.
For large $M$ it approaches to unity for both signals 
$\tilde{s}_n$ and $s'_n$ at $\epsilon =0$.
However, for a small number of measurements
$(5 \leq M \leq 50)$ the fidelity is significantly higher for
measurements performed in the frequency domain after QFT (Fig. 3 inset).
The presence of noise in the quantum gates used in QFT for
$s'_n$ reduces the value of $\tilde{f}$ but
for $5 \leq M \leq 50$ and $\epsilon \leq 0.1$ this reduction
is not significant. The drop of $\tilde{f}$  
becomes considerable only at relatively large amplitudes
with $\epsilon > 0.2$ as it is shown in Fig. 4.
The residual level of $\tilde{f}$ at maximal $\epsilon \approx 1$
is in agreement with the statistical estimate
according to which 
$\tilde{f}(\epsilon =1) \approx \sqrt{n_i/2^{n_f}} \approx 0.2$,
where $n_i$ is the number of frequencies per frame for
the original signal ($n_i \approx 20$ according to Fig. 2).
At small $\epsilon$ the drop of $\tilde{f}$ is quadratic in $\epsilon$
($1-\tilde{f} \sim \epsilon^2 n_f^2$) since 
each gate transfers about of $\epsilon^2$ amount of probability
from ideal computational state to all other states \cite{levi}.
The obtained results show that the MP3-like strategy
adapted to the quantum signals allows to recover
human speech  with a significantly smaller 
number of measurements with a reduction factor of 10-20.  

\begin{figure}[t!]
\epsfxsize=3.2in
\epsffile{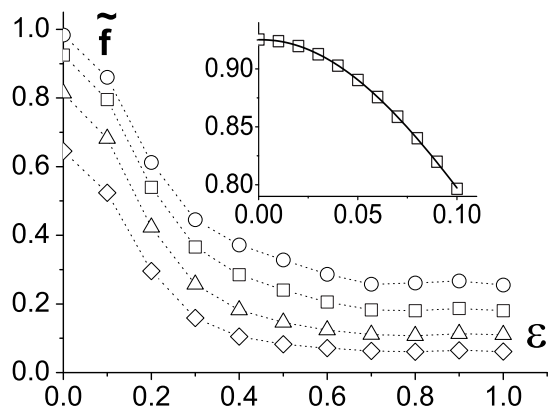}
\vglue -0.8cm
\caption{Dependence of fidelity $\tilde{f}$ for the sound signal $s'_n$
on the noise amplitude $\epsilon$ for the number of measurements 
$M=1000$ (o), 100 (squares),
20 (triangles), 5 (diamonds). Inset shows a data fit 
$1-\tilde{f} =0.36 \epsilon^2 n_f^2$
(full curve) at small $\epsilon$,  $M=100$.
}
\label{fig4}       
\end{figure}

Above we assumed that the sound signal is already 
encoded in the wavefunction.
For certain quantum objects such an encoding can be done 
efficiently.  As an example we consider the wavefunction evolution
described by the quantum sawtooth map
\begin{equation}
\overline{\psi}=\hat{U}\psi =
e^{-iT\hat{l}^2/2}
e^{ik\hat{\theta}^2/2}\psi,  
\label{qumap}
\end{equation}
where $\hat{l}=-i\partial/\partial\theta$, $\hbar=1$,
$k, T$ are dimensionless map parameter and $\overline{\psi}$
is the value of $\psi$ after one map iteration (we set $\hbar=1$).
In the semiclassical limit $k \gg 1$, $T \ll 1$ the chaos
parameter of the model is $K=kT=const$.
The efficient quantum algorithm for the simulation
of this complex dynamics was developed and tested in \cite{kr,benenti}.
The computation is done for the wavefunction $\psi$ on a discrete
grid with $N=2^{n_q}$ points with
$\theta_n=2\pi n/N, \; n=1,...,N$ in $\theta$-representation
and $l+N/2=1,..,N$ in momentum representation.
Here, as before $n_q$ is the number of qubits
in a quantum computer and in
$\theta$-representation $\psi= \sum_n \psi(\theta_n) |n>$ 
is encoded in the register $|n> = |a_1...a_i...a_{n_q}>$.
The transition between $n$ and $\theta$ representations is done by QFT
and one map iteration is computed in $O(n_q^2)$ quantum gates
for an exponentially large vector of size $2^{n_q}$ \cite{benenti}.
To study the sound of quantum wavefunctions of map (\ref{qumap})
we choose here a case with $K=-0.5$, $T=2\pi/N$ and $n_q=14$
corresponding to a complex phase space structure.

The signal encoded in the wavefunction $\psi(\theta_n)$
after $t$ map iterations 
can be treated in a way similar to one used before
for the HAL speech $s_n$.  The measurements in $\theta$-basis
give the signal $\tilde{s}_n= |\psi(\theta_n)|$ which
however requires a large number of them to suppress noise
(also the phase is completely lost). Another method works as
for $s'_n$ signal: first QFT is performed 
on $n_f=5$ less significant qubits giving
$\psi = \sum_{k,j} S_{k,j}|k,j>$ and then
the measurements  are done to determine the
instantaneous spectrum amplitudes $|S_{k,j}|$ of $\psi(\theta_n)$
(here $k=1,...,2^9$ is the frame number,
$j=1,...,2^5$ is the index of frequency harmonics
and $\Delta n = 2^5$).
The sound of quantum wavefunction is recovered
via the inverse classical FFT giving
$s'_n$ signal defined before. Examples of restored sound 
are given at \cite{qaudiosite} and clearly show that the quality
of MP3-like signal $s'_n$ is much higher compared to $\tilde{s}_n$
(we use sampling rate $f=1$ kHz for this case with $n_q=14$).

\begin{figure}[t!]
\epsfxsize=3.2in
\epsffile{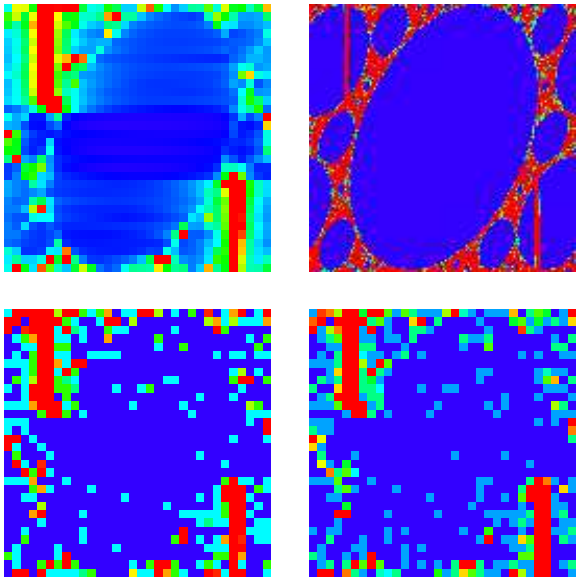}
\vglue -0.5cm
\caption{(color) Left: coarse grained diagram of $S^{(g)}$
for the sound signal $s'_n$ obtained from 
the quantum computation (\ref{qumap}) of
the wavefunction after $t=100$ map iterations, 
top panel shows exact distribution
$S^{(g)}$ and bottom panel is for a  number of measurements
as in Figs. 1,2. Bottom right: same as
bottom left but with noise amplitude $\epsilon = 0.05$
in the quantum gates. Right top: the exact Husimi
distribution $h(l,\theta)$. The initial state is a momentum eigenstate
at $l=100-N/2$. The color and axes are as in Fig. 2.
}
\label{fig5}       
\end{figure}

A more detailed analysis of the quantum sound $s'_n$ 
can be obtained from the coarse grained spectrum digram
similar to the one in Fig. 2. The coarse graining of 
$S_{k,j}$ is done by measuring  5 most significant
and 5 less significant qubits corresponding to
$a_1 a_2 ... a_5$ and $a_{10} a_{11} ... a_{14}$
that gives coarse grained distribution $S^{(g)}$ 
in $32 \times 32$ cells.
The diagrams of $S^{(g)}$ obtained for infinite and finite
number of measurements are displayed in Fig. 5
(left top and bottom respectively). The exact 
diagram shows an interesting structure which is recovered
with a finite number of measurements. This structure remains
robust against noise in the quantum gates
used for computation of $100$ map iterations 
and final QFT (Fig. 5 right bottom). 

The origin of this structure becomes clear after its
comparison with the coarse grained Wigner function
called the  Husimi distribution $h(\theta,l)$ \cite{levi,husimi} 
which is shown in Fig. 5 (right top) which is very close to
$S^{(g)}$ (left top). Indeed, $h(\theta,l)$ is defined 
in the phase space $(l,\theta)$ by 
\begin{equation}
h(l,\theta)= \sum_{l'=l-N/2}^{l+N/2} G(l'-l) \psi(l') e^{i l' \theta}
\label{husimi}
\end{equation}
where the gaussian smoothing function is 
$G(l'-l) =(T/\pi)^{1/4} e^{-T(l'-l)^2/2} /\sqrt{N}$ \cite{levi,husimi}.
The Husimi distribution is always positive and
gives a direct comparison between the classical phase space
 Liouville density distribution
and a quantum wavefunction. In fact the coarse grained distribution
$S^{(g)}$ is also given by equation (\ref{husimi})
where $G(l')$ is replaced by a constant in the interval $\Delta l' =2^{5}$
and zero outside that corresponds to the application of QFT to less
significant qubits $n_f=5$. Such a replacement modifies
the values of coarse grained $h(\theta,l)$
but this modification remains small if $\Delta l' \gg 1$
\cite{frahm}. 
As a result we may argue that the signal $s'_n$
represents the quantum sound of coarse grained Wigner function.

In conclusion, our results show that sound signals stored in a
quantum memory can be reliably recognized and recovered 
on realistic quantum computers. The method proposed 
allows to obtain  sound of quantum wavefunctions
that  can be useful for future quantum telecommunications.

This work was supported in part by the EC IST-FET project
EDIQIP and the NSA and ARDA under ARO contract No. DAAD19-01-1-0553. 


\begin{thebibliography}{99}
\bibitem{mp3} http://www.mpeg.org/MPEG/audio.html
\bibitem{chuang}  M.A.~Nielsen and I.L.~Chuang {\it Quantum Computation and 
       Quantum Information}, Cambridge Univ. Press, Cambridge (2000).
\bibitem{qcom}   N.~Gisin, G.~Ribordy, W.~Tittel, and H.~Zbinden,
                 Rev. Mod. Phys. {\bf 74}, 145 (2002).
\bibitem{shor}  P.W.Shor, in {\it Proc. 35th Annual Symposium
       on Foundation of Computer Science}, Ed. S.Goldwasser (IEEE Computer
       Society, Los Alamitos, CA, 1994), p.124.
\bibitem{cory}Y.S.~Weinstein, M.A.~Pravia, E.M.~Fortunato,
                 S.~Lloyd, and D.G.~Cory, Phys. Rev. Lett. 
                 {\bf 86}, 1889 (2001).
\bibitem{chuang1}  L.M.K.Vandersypen, M. Steffen, G. Breyta,
                 C.S. Yannoni, M.H. Sherwood, and I.L. Chuang,
                 Nature {\bf 414}, 883 (2001).
\bibitem{blatt}  S.Gulde, M.Riebe, G.P.T.Lancaster, C.Becher, J.Eschner, 
      H.H\"affner, F.Schmidt-Kaler, I.L.Chuang and R.Blatt, Nature {\bf 421},
      48 (2003).
\bibitem{wigner}  E.~Wigner Phys. Rev. {\bf 40}, 749 (1932);
            M.~V.~Berry, Phil. Trans. Royal Soc. {\bf 287}, 237 (1977).
\bibitem{laflamme}  C.~Miquel, J.~P.~Paz, M.~Saraceno, E.~Knill, 
                   R.~Laflamme and C.~Negrevergne, Nature {\bf 418}, 59 (2002).
\bibitem{levi} B.~L\'evi, B.~Georgeot and D.L.~Shepelyansky,
               Phys. Rev. E {\bf 67}, 046220 (2003).
\bibitem{hal} Sound is available at http://www.palantir.net/cgi-bin/file.cgi?file=wav/hal9000.wav
\bibitem{qaudiosite} http://www.quantware.ups-tlse.fr/qaudio/
\bibitem{kr} B.~Georgeot and D.~L.~Shepelyansky, Phys. Rev. Lett. {\bf 86}, 2890
  (2001). 
\bibitem{benenti} G.~Benenti, G.~Casati, S.~Montangero and D.~L.~Shepelyansky, 
  Phys. Rev. Lett. {\bf 87}, 227901 (2001).
\bibitem{husimi} S.-J. Chang and K.-J. Shi, Phys. Rev. A 
            {\bf 34}, 7 (1986).
\bibitem{frahm} Detailed analysis of quantum computation of
        Husimi distribution is done by K.M.~Frahm (in preparation, 2003).
 
\end{thebibliography}
\end{document}